# Electronic stopping power of hydrogen in a high-k material at the stopping maximum and below

D. Primetzhofer[*]

*Ion physics, Department of Physics and Astronomy, Uppsala Universitet, Box 516, S-751 20 Uppsala, Sweden*

**Abstract**

Electronic energy loss of hydrogen ions in $HfO_2$ was investigated in a wide energy range in the medium and low energy ion scattering regime. Experiments by Time-Of-Flight Medium-Energy Ion Scattering (TOF-MEIS) with proton and deuteron projectiles were performed in backscattering geometry for nm-films of $HfO_2$ on Si with an ultrathin $SiO_2$ interface layer prepared by ALD. At energies around the stopping maximum excellent agreement is found with earlier results from Behar et al. (Phys. Rev. A 80 (2009) 062901) and theoretical predictions. Towards lower energies discrepancies between experiment and calculations increase slightly. The low energy data exhibits excellent velocity proportionality and indicates the absence of clear effects due to distinct electronic states. Thus, also no apparent velocity threshold can be predicted within the experiments uncertainty from the present data. The magnitude of the energy loss is discussed in terms of a simple free-electron model and compared with the expected electron densities from plasmon frequencies.

*) Corresponding author: daniel.primetzhofer@physics.uu.se.at , Tel.: ++46/18 471 3040

## Introduction:

The deceleration of ions in solids due to interaction with the electrons of the target atoms has been of interest since the early days of atomic physics [1]. The specific energy loss is also the key quantity when the path length travelled by an ion in a solid is of relevance. Thus, the energy exchanged between an intruding ion and the electronic system of the target is not only of fundamental interest but of uttermost importance for depth profiling techniques employing ion beams and applied to thin films as well as ion beam based material modification.

The decelerating force dE/dx, i.e. the mean energy loss per path length, an ion experiences in a material is commonly denoted as the stopping power S of the material [2]. Since this concept introduces a dependence on the number density n of the penetrated solid it can be more convenient to make use of the electronic stopping cross section (SCS) $\varepsilon$ defined via $\varepsilon = (1/n) \cdot dE/dx$. Consequently the SCS has the dimension of an energy times a unit area.

For ions in the MeV energy regime, as employed in classical ion beam analysis (IBA) a fundamental theoretical description of the energy loss due to interaction with electrons is given by Bethe and Bloch [3],[4],[5] and subsequent improvements [6],[7],[8],[9] which yield both qualitative and quantitative agreement at sufficiently high energies. When decreasing the ion energy, a concept of effective charge has to be introduced to allow for a reasonable prediction of the electronic energy loss [10],[11],[12] since the ions may have bound electrons. Eventually, the stopping power exhibits a pronounced maximum, with the energy dependent on the ion species (e.g. around 100 keV for protons [13]) due to the interplay of decreasing effective charge, increasing scattering cross sections and decreasing maximum energy transfer.

This understanding of the energy loss processes has laid the foundation for the success of IBA using MeV ions. At these energies a good qualitative and quantitative prediction of S for almost any compound is possible [14],[15],[16],[17] and large compilations of experimental and theoretical data for many different materials are available [18].

However, in consequence of the ongoing miniaturization in electronics and sensor technology classical IBA has reached its limits in depth resolution. By the use of different detectors and reduce the ion energy, and thus the relative energy loss per unit path length this constraint can be overcome. As examples, Medium-Energy Ion Scattering (MEIS) or High-Resolution Rutherford Backscattering Spectrometry (HR-RBS) making use of magnetic and electrostatic spectrometers (see e.g. [19],[20]) as well as modified solid state detectors [21] were developed. Some of these approaches yield monolayer resolution [22].

Along with the decrease in ion energy a more urgent demand for a similar accurate description of energy loss at energies around and below the stopping power maximum is created. The simplest model system imaginable is a free electron gas (FEG) for which S is predicted to be proportional to the ion velocity below the stopping maximum [23]. Velocity proportional energy loss has in fact been observed for most metals since in metals occupied electronic states extend up to the Fermi level $E_F$ and the unoccupied states available above $E_F$ allow for in principle arbitrarily small energy transfers. Only for noble metals and very low energies, i.e. below 10 keV for protons or ~25 keV for He ions, excitation thresholds are observed, typically associated with the finite binding energy of d-electrons [24],[25],[26].

Since insulators and semiconductors both exhibit a region of forbidden states in their electronic band structure [27] and their conduction band is unoccupied, in order to excite an electron-hole pair, a minimum excitation energy is necessary to permit an electron to overcome the band gap. This finite minimum excitation energy on the electronic stopping power was expected to lead to strong nonlinearities in e observable at ion energies of several keV [23]. Instead, however, in most systems velocity proportional stopping was observed even for velocities below the binary collision threshold [28],[29],[30], with the exception of protons in He, for which a strong deviation from $\varepsilon \sim v$ was observed [31]. Eventually, thresholds for electronic stopping were found [32],[33],[34], which were, however, restricted to the sub-keV energy regime, and thus not of real relevance for ion beam analysis at medium energies. Still, the absolute magnitude of electronic stopping in these compounds was in some cases found to be strongly affected by the chemical composition and clear effects on the absolute magnitude of electronic stopping are observed at energies of about 100 keV and protons [35],[36]. For He and similar energies additional effects due to charge changing collisions can alter the observed energy loss and complicate theoretical description [37].

In the present study electronic energy loss of hydrogen ions in $HfO_2$ was investigated in a wide energy range (5 – 100 keV). These experiments are of interest for two reasons: From a fundamental viewpoint this system deserves special interest since $HfO_2$ is an insulator with a large band gap of about 6 eV and with distinct electronic states due to the 4f-electrons of Hf forming a narrow sub-band around -17 eV and the oxygen 2p-electrons around -5 eV [38],[39].. Due to the strongly covalent character of the bonds a deviation from stopping power values predicted from the elements of the compound according to Bragg's rule can be expected at energies around the stopping maximum [40].

From an applied point of view, stacks containing $HfO_2$ form a key ingredient in nowadays semiconductor technology, both in research and industry. A large number of studies which make use of medium energy ions to characterize stacks of high-k materials like $HfO_2$ has been published in the last years [41],[42],[43],[44]. Thus accurate reference stopping powers for this compound are of relevance for obtaining correct depth scales in ion beam based analysis of the above mentioned systems. Interestingly, stopping power data for $HfO_2$ is scarce, with some data available at energies above 100 keV [45] but lacking information towards lower energies [18].

**Experiment and evaluation**

Sample preparation was done ex-situ. The $SiO_2$ was grown via $O_3$ oxidization and the $HfO_2$ by atomic layer deposition (ALD). A first thickness estimation for the samples was obtained by ellipsometry measurements. Calibration experiments using Rutherford Backscattering Spectrometry (RBS) were performed employing a 5 MV 15SDH-2 tandem accelerator at the Ångström laboratory at Uppsala University. An implanter-based TOF-MEIS set up at Uppsala University [46] was used for the stopping power measurements at medium ion energy. Only the most important features of the set-up will be summarized here: The set-up is capable to produce atomic and molecular beams of many different ion species in an energy range from approximately 20 – 180 keV. A moveable large solid-angle (diameter of the active area 120 mm) microchannelplate (MCP) detector permits to record position sensitive time-spectra for different scattering angles in backscattering geometry. Due to the large area of the detector the particle flux can be kept extremely low (on the order of 10 nC per spectrum). This permits to use also heavier ions than hydrogen at low ion energies without any significant sample damage due to e.g. sputtering.

Backscattering spectra were recorded in both set-ups, for 2 MeV $He^+$ ions to obtain the Hf concentration in atoms/cm2, and by $H^+$, $H_2^+$, $D^+$ and $D_2^+$ ions with primary energies $E_0$ between 20 and 100 keV which yields an effective energy range of 5 – 100 keV/u. for the deduced stopping power data. Two examples for energy-converted TOF-MEIS spectra obtained in this study are shown in Fig. 1 a) & b).

In both spectra the dominant feature is due to backscattering from Hf in the $HfO_2$ film. In their main features, the spectra still resemble typical RBS data for thin films with some influence of multiple scattering being noticeable in a low energy tail of the spectrum (more pronounced for 50 keV $D_2^+$ corresponding to 25 keV D or 12.5 keV H in the energy loss, respectively). Due to the excellent energy resolution, small inhomogeneities in the film thickness (~10%) can be deduced from the spectrum obtained for 90 keV $H_2^+$.

To deduce information on the inelastic energy loss, the backscattered yield for a certain narrow scattering angle interval is simulated by Monte-Carlo trajectory calculations by the TRBS-code (TRim for BackScattering) [47] (red lines in Figs. 1a) & b)). In the simulations, a correction factor for electronic stopping and different interatomic potentials with tuneable screening (ZBL, TFM [48],[49]) are available. The electronic energy loss can be tuned by a constant scaling factor and is modeled according to [48]. For a fixed thickness of the $HfO_2$ layer this factor is adjusted until the width of the feature from backscattering from Hf is matched. For thin targets with $\Delta E \ll E_0$ this permits an accurate evaluation of the experienced electronic energy loss, virtually independent of the stopping model in the program. There are several expected contributions to the error of the deduced stopping cross sections. One error source is the thickness calibration by RBS, where the stopping power in the Si substrate is entering the evaluation of the Hf concentration. This introduces a possible systematic error of about 2.5% to the final results. Experiments statistics, as well as the accuracy of the fitting procedure to the leading and the trailing edge of the experimental spectra is expected to yield an error well below 5%. One remaining uncertainty relevant in stopping experiments for compound materials is the influence of the stoichiometry. Poor film quality and a fit exclusively to the features from scattering from Hf and Si leave the possibility to obtain systematically wrong stopping cross sections, if the investigated compound actually would not have been $HfO_2$. Although the growth of $HfO_2$ layers is a standard process [50] and the joint results from ellipsometry, capacitance measurements and ion scattering suggest excellent film stoichiometry, a possible systematic error of at most 5% may be attributed to this issue. A more thorough discussion of this possible problem, by analysis of spectra obtained for backscattering of He ions, for which this question is more relevant due to the low energy of projectiles backscattered from O, can be found elsewhere [51]. Considering all these effects the cumulative error of the deduced stopping cross sections is expected to not exceed 7% with highest uncertainty at lowest energies.

**Results and discussion**

The obtained experimental results for the stopping cross section $\varepsilon$ of hydrogen in $HfO_2$ are shown in Fig. 2 (red asterisks). The experimental data features a low scatter, and no isotope effects or vicinage effects due to fragmentation of molecular ions are observed within the experiments precision, in accordance with expectations [52]. Also shown is the previous experimental data from Behar et al. [45] (open squares) as well as their calculations based on the Mermin energy loss function – generalized oscillator strength (MELF-GOS) model [53],[54] both for the calculated equilibrium charge state as well as for the charge state of the projectile fixed to +1 (red dash double-dotted and violet dash dotted line, respectively). The figure also shows a calculation using SRIM2011 (black solid line). Note, that the experimental data from the two investigations fit excellently at the stopping maximum. Predictions by the MELF-GOS model are found too low in the whole regime of investigated velocities. Note that the calculations fit better for bare protons that for the calculated equilibrium charge state. SRIM slightly overestimates experimental data in the whole experimentally investigated region which is in accordance with the expectations of a violation of Bragg's rule at low velocities due to chemical effects. A best fit according to [48] is also shown as a dashed line through both experimental datasets in the figure.

For comparison with theory and other datasets for low ion energies the same dataset is also presented as a function of the ion velocity in Fig. 3. As can be readily seen from the figure, data exhibits excellent velocity proportionality at velocities below 1 a.u. From the present data, no evidence for a deviation from this velocity proportionality towards lower velocities can be found within the experiments statistics. This indicates, that the contribution of the f-electrons to electronic stopping at these energies is negligible, which is also in accordance with earlier investigations for protons in Zn, for which the d-electrons with a binding energy of 8 eV also were not found to lead to a noticeable deviation from $\varepsilon \sim v$ at low energies [55]. Thus electronic stopping at $v < 1$ a.u. is expected to be mainly due to the O 2p-electrons, which form the main contribution to the density of states at the high-energy edge of the valence band [38]. One can perform a naive calculation based on a free-electron gas model for the electronic energy loss [56]: Whereas from the electronic structure up to 12 O 2p-electrons per molecule have to be expected, a calculation from the observed proportionality coefficient of e at $v < 1$ a.u results in an effective number of electrons of only ~3.7 e$^-$ per molecule. This density would correspond to a plasmon energy of 12 eV different from experimental observations of about 15-16 eV [57] whereas contribution of 12 electrons would lead to an energy of 22 eV. This result is in agreement with what has been observed previously for electronic states with excitation thresholds in noble metals (d-electrons), where calculations from the expected electron density also clearly overestimated the experimentally observed energy loss.

An extrapolation towards $\varepsilon = 0$ from data at velocities below 1 a.u. in the ion velocity yields a negative threshold of -0.03 +/- 0.17 a.u. Thus, also a possible small (positive) threshold velocity cannot be excluded even if the data would scale linearly with the ion velocity down to $v = 0$. For comparison, a linear fit for a threshold velocity of 0.4 a.u. is depicted as a dashed line in Fig. 3. This specific choice for the threshold velocity is made based on the naive assumption that the threshold scales approximately linearly with the band gap width, an observation which is indeed made for H and He in LiF and KCl, but instead has not been verified for other systems like Ge [58] for which an unexpectedly large threshold velocity was observed. In any case, as has been observed for protons in LiF only a minor change in slope at velocities below 0.4 a.u. from the present velocity proportionality is sufficient to lead to a threshold-like behavior [59].

**Summary and conclusions:**

The electronic stopping cross section of $HfO_2$ for hydrogen ions has been measured experimentally for equivalent proton energies in the range from 5 to 100 keV. Data fits excellent to data previously measured at higher energies by Behar et al. with a pronounced stopping power maximum observed at 100 keV/amu. For energies below 25 keV, i.e. 1 a.u. in the ion velocity, data exhibits excellent velocity proportionality.

This proportionality is an indication for the absence of distinct excitation limits of certain electronic states in the investigated energy range. The magnitude of the observed low-energy stopping is in accordance with expectations for a free electron gas with an effective density as deduced from the $HfO_2$ plasmon energy. SRIM predictions are found to only slightly overestimate stopping cross sections in the MEIS energy regime, relevant for the commonly performed MEIS analytics of high-k stacks. Discrepancies with the MELF-GOS model are found to increase towards lower energies. Further theoretical investigation of the electronic interactions at energies below 100 keV would be desirable in order to identify the electronic states contributing to ion deceleration. From a fundamental point of view, also further experimental investigations of $HfO_2$ at even lower ion energies would be of interest in order to search for effects due to distinct electronic states of the material.

**Acknowledgements**

**References**

[1] J.J. Thomson, Phil. Mag. 23 (1912) 449

[2] P. Sigmund, ICRU News 1, (2000), 17

[3] H. Bethe, Ann. d. Physik, 5 (1930) 325

[4] H. Bethe and J. Ashkin, Experimental Nuclear Physics, ed. E. Segré, J. Wiley, New York, (1953) 253

[5] F. Bloch, Annalen der Physik 16 (1933) 285

[6] H. Bethe, Z. Phys. 76 (1932) 293

[7] Andersen H.H., Simonsen H., Sørensen H.: Nuclear Physics A125 (1969) 171

[8] W. H. Barkas, J. N. Dyer and H. H. Heckmann: Phys. Rev. Lett. 11 (1963) 26

[9] U. Fano, Annu. Rev. Nucl. Sci. 13, (1963) 1.

[10] W. Brandt and M. Kitagawa, Phys. Rev. B 25 (1982) 5631

[11] A. Arnau, M. Peñalba, P. M. Echenique, and F. Flores, Nucl. Instr. Meth. B 69 (1992) 102

[12] N.R. Arista, K. Dan. Vidensk. Selsk. Mat. Fys. Medd. 52, (2006) 595

[13] C.C. Montanari, J. E. Miraglia, and N. R. Arista, Phys. Rev. A 66 (2002) 042902

[14] ASTAR: Stopping power and range tables for helium ions available from: http://physics.nist.gov/PhysRefData/Star/Text/ASTAR.html

[15] PSTAR Stopping power and range tables for protons available from: http://physics.nist.gov/PhysRefData/Star/Text/PSTAR.html

[16] SRIM, J.F. Ziegler, available from: http://www.srim.org/

[17] J.F. Ziegler, Nucl. Instr. and Meth. B 219 (2004) 1027

[18] H. Paul: Stopping powers for light ions: available from: http://www.exphys.jku.at/stopping/

[19] J. Vrijmoeth, P.M. Zagwijn, J.W.M. Frenken, J.F. van der Veen, Phys. Rev. Lett. 67 (1991) 1134

[20] K. Kimura, K. Ohshima, M. Mannami, Appl. Phys. Lett. 64 (1994) 2232

[21] M. Geretschläger, Nucl. Instr. and Meth. B 204 (1983) 479

[22] Y. Kido, T. Nishimura, Y. Hoshino, H. Namba, Nucl. Instrum. Methods B 161–163 (2000) 371

[23] E. Fermi, E. Teller, Phys. Rev. 72 (1947) 399

[24] J.E. Valdes, J.C. Eckardt, G.H. Lantschner, and N.R. Arista, Phys. Rev. A 49 (1994) 1083

[25] E. A. Figueroa, E. D. Cantero, J. C. Eckardt, G. H. Lantschner, J. E. Valdés, and N. R. Arista, Phys. Rev. A 75 (2007) 010901

[26] S.N. Markin, D. Primetzhofer, M. Spitz, and P. Bauer, Phys. Rev. B (2009) 205105

[27] H. Ibach, O. Lüth, Festkörperphysik: Einführung in die Grundlagen, Springer Verlag, 4. Auflage (1988)

[28] K. Eder, D. Semrad, P. Bauer, R. Golser, P. Maier-Komor, F. Aumayr, M. Peñalba, A. Arnau, J. M. Ugalde, and P. M. Echenique, Phys. Rev. Lett. 79 (1997) 4112

[29] J. I. Juaristi, C. Auth, H. Winter, A. Arnau, K. Eder, D. Semrad, F. Aumayr, P. Bauer, and P. M. Echenique, Phys. Rev. Lett. 84 (2000) 2124

[30] S.P. Møller, A. Csete, T. Ichioka, H. Knudsen, U.I. Uggerhøj, and H.H. Andersen, Phys. Rev. Lett. 93, 042502 (2004).

[31] R. Golser, D. Semrad, Phys. Rev. Lett. 66 (1991) 1831

[32] H.O. Funsten, S.M. Ritzau, R.W. Harper, J.E. Borovsky and R.E. Johnson, Phys. Rev. Lett. 92, 213201 (2004).

[33] L. N. Serkovic, E. A. Sánchez, O. Grizzi, J. C. Eckardt, G. H. Lantschner, and N. R. Arista, Phys. Rev. A 76 (2007) 040901

[34] S.N. Markin, D. Primetzhofer, and P. Bauer, Phys. Rev. Lett 103 (2009) 113201

[35] D.I. Thwaites, Radiat. Res. 95 (1983) 495

[36] D. Primetzhofer, S.N. Markin, P. Bauer, Nucl. Instr. Meth. B 269 (2011) 2063

[37] D. Primetzhofer, Phys. Rev. B 86 (2012) 094102

[38] T. V. Perevalov, V. A. Gritsenko, S. B. Erenburg, A. M. Badalyan, Hei Wong, C.W. Kim, J. Appl. Phys. 101 (2007) 053701

[39] Suzer, S. Sayan, M.M. Banaszak Holl, E. Garfunkel, Z. Hussain, N.M. Hamdan, J. Vac. Sci. Technol., A 21 (2003) 106

[40] W. H. Bragg and M. A. Elder, Philos. Mag., 10, 318 (1905)

[41] Y. Hoshino, Y. Kido, K. Yamamoto, S. Hayashi and M. Niwa, Appl. Phys. Lett. 81 (2002) 2650

[42] K. Saskawa, K. Fujikawa, T. Toyoda, Appl. Surf. Sci. 255 (2008) 1551

[43] K. Nakajima, S. Joumori, M. Suzuki, K. Kimura, T. Osipowicz, K. L. Tok, J. Z. Zheng, A. See, and B. C. Zhang, Appl. Surf. Sci. 237, 416 (2004)

[44] H. S. Chang, H. Hwang, M. H. Cho, and D. W. Moon, Appl. Phys. Lett. 86 , 031906 (2005)

[45] M. Behar, R.C. Fadanelli, I. Abril, R. Garcia-Molina, C.D. Denton, L.C.C.M. Nagamine and N.R. Arista, Phys. Rev. A 80 (2009) 062901

[46] M. K. Linnarsson, A. Hallén, J. Åström, D. Primetzhofer, S. Legendre, and G. Possnert, Rev. Sci. Instr. 83 (2012) 095107

[47] J.P. Biersack, E. Steinbauer, P. Bauer, Nucl. Instr. and Meth. B 61 (1991) 77

[48] J.F. Ziegler, J.P. Biersack, U. Littmark, The Stopping and Range of Ions in Solids, Vol. 1, Pergamon, New York, 1985

[49] G. Molière, Z. Naturforsch. 2a (1947) 133.

[50] E. Dentoni Litta, P.-E. Hellstrom, C. Henkel, M. Ostling, Proc. of the 13th Intern. Conf. on Ultimate Integration on Silicon (ULIS), (2012) 105

[51] D. Primetzhofer, E. Dentino Litta, A. Hallén, M. Linnarsson, G. Possnert, subm. To Nucl. Instr. Meth. B

[52] S. M. Shubeita, P. L. Grande, J. F. Dias, R. Garcia-Molina, C. D. Denton and I. Abril, Phys. Rev. B 83 (2011) 245423

[53] I. Abril, R. Garcia-Molina, C. D. Denton, F. J. Pérez-Pérez, and N. R. Arista, Phys. Rev. A 58, 357 (1998).

[54] S. Heredia-Avalos, R. Garcia-Molina, J. M. Fernández-Varea, and I. Abril, Phys. Rev. A 72, 052902 (2005).

[55] G. Martinez-Tamayo, J.C. Eckardt, G.H. Lantschner, and N.R. Arista, Phys. Rev. A 51(1995) 2285.

[56] I. Nagy, A. Arnau, and P. M. Echenique, Phys. Rev. A 40 (1990) 987

[57] X. F. Wang, Quan Li, R. F. Egerton, P. F. Lee, J. Y. Dai, Z. F. Hou, and X. G. Gong, J. Appl. Phys. 101, 013514 (2007)

[58] D. Roth, D. Goebl, D. Primetzhofer, P. Bauer, Nucl. Instr. Meth. B in press

[59] M. Draxler, S. P. Chenakin, S. N. Markin, and P. Bauer, Phys. Rev. Lett. 95, 113201 (2005).

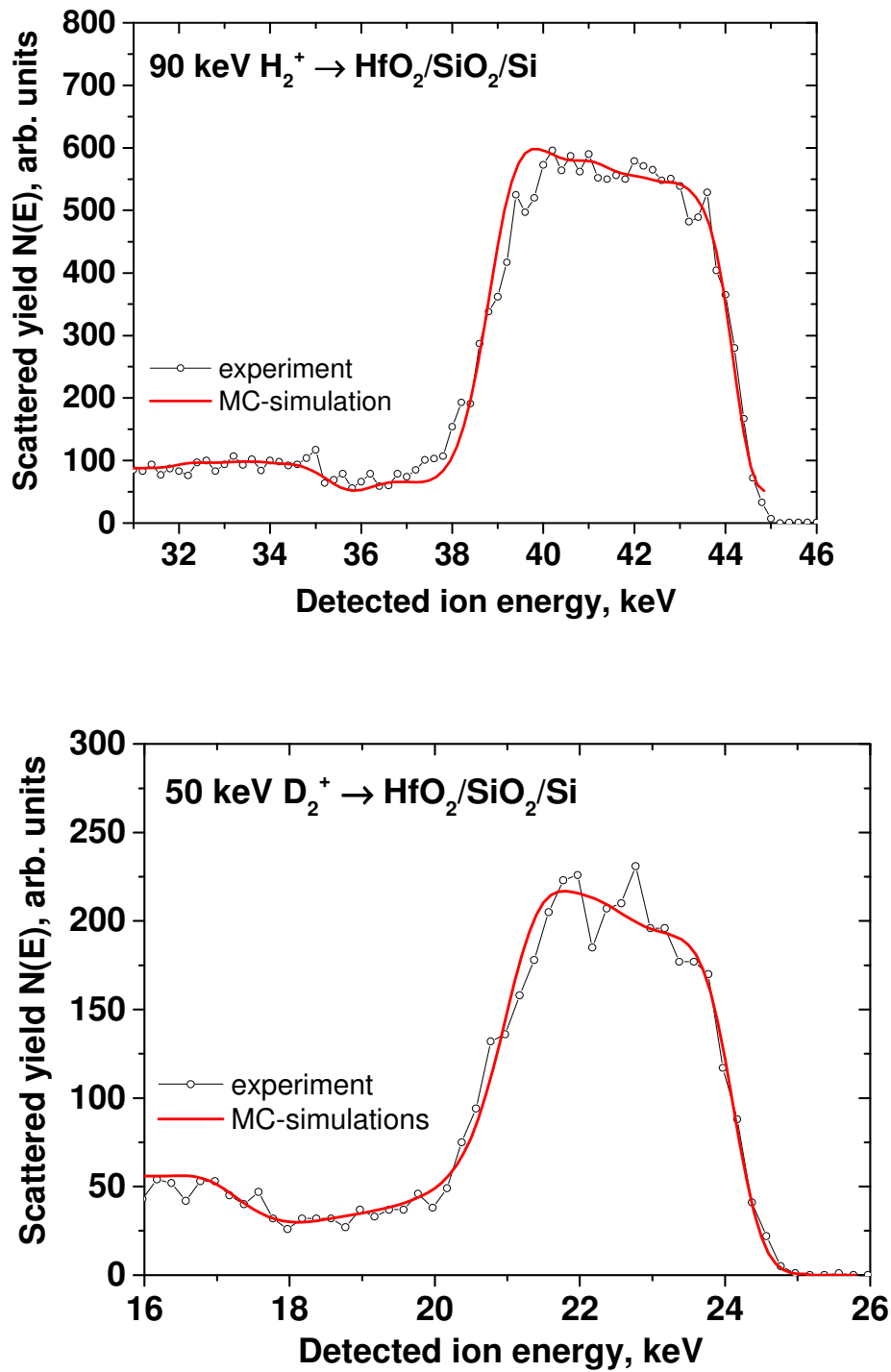

Fig 1: Experimental energy-converted TOF-MEIS spectra obtained for a) 90 keV $H_2^+$ ions b) 50 keV $D_2^+$ ions. Also shown are Monte-Carlo simulations using TRBS [47] which permits to obtain the electronic stopping cross section $\varepsilon$ from a fit to the spectrum width. Note that the excellent energy resolution in a) permits to obtain information on the small (~10%) inhomogeneity of the film. For more details, see text.

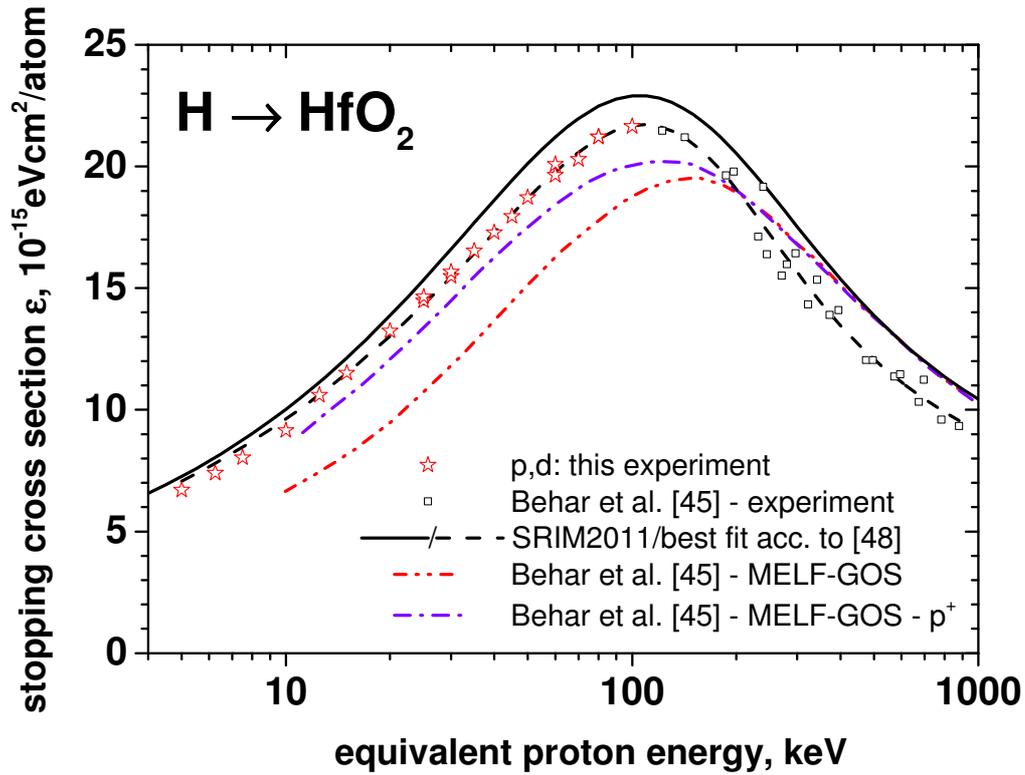

Fig.2: Experimentally obtained stopping cross sections for protons in $HfO_2$ (red asterisks) as a function of projectile velocity. Also shown is previously measured data from Behar et al. [45] at elevated energies (open squares) and theoretical calculations from the same work (dash-dotted and dash double-dotted lines) as well as SRIM predictions [16] (full line). For details see text.

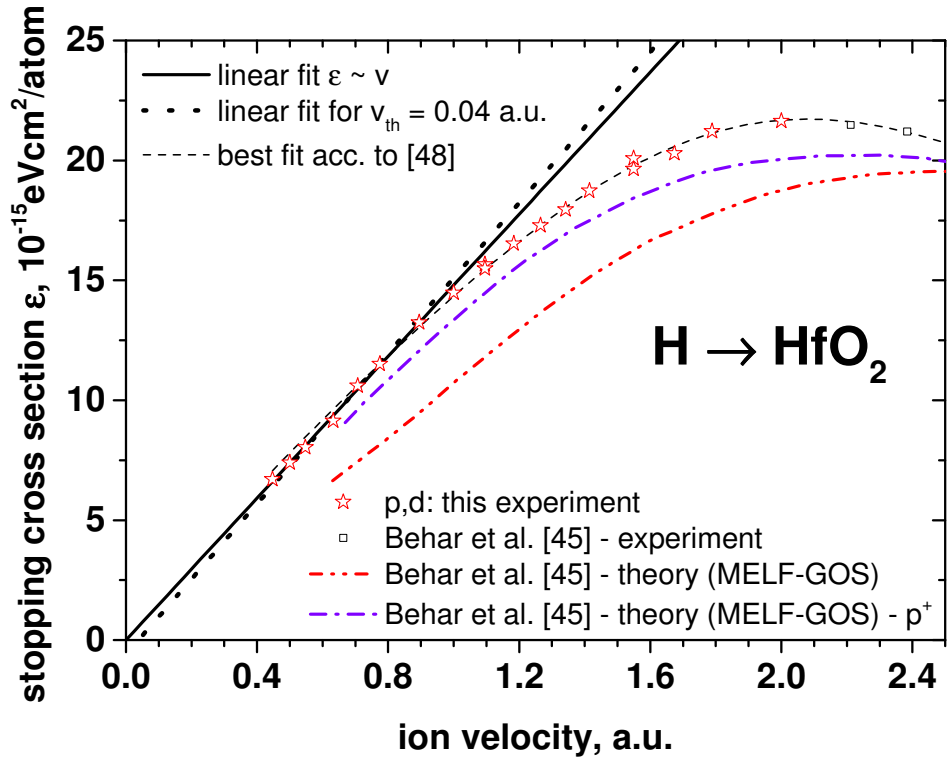

Fig. 3: Experimentally obtained stopping cross sections of protons in $HfO_2$ as a function of the projectile velocity (same data as in Fig.2). For details see text.